\documentclass[twocolumn,aps,amsmath,showpacs]{revtex4}
\usepackage{graphicx}
\newcommand{\be}{\begin{equation}}
\newcommand{\ee}{\end{equation}}

\newcommand{\bea}{\begin{eqnarray}}
\newcommand{\eea}{\end{eqnarray}}
\newcommand{\bd}{\begin{displaymath}}
\newcommand{\ed}{\end{displaymath}}
\newcommand{\bi}{\begin{itemize}}
\newcommand{\ei}{\end{itemize}}
\newcommand{\bc}{\begin{center}}
\newcommand{\ec}{\end{center}}
\newcommand{\bfl}{\begin{flushleft}}
\newcommand{\efl}{\end{flushleft}}
\newcommand{\bfr}{\begin{flushright}}
\newcommand{\efr}{\end{flushright}}
\newcommand{\f}{\frac}

\def\bk{{\bf k}}

\def\6{\partial} \def\a{\alpha} 
  \def\ve{\varepsilon}

\def\={\!\!\!&=&\!\!\!}
\def\+{\!\!\!&&\!\!\!+~}
\def\-{\!\!\!&&\!\!\!-~}

\begin{document}
\author{I. \c{T}ifrea$^{1}$, G. Pal$^2$, and M. Crisan$^{3}$}

\affiliation{$^1$Department of Physics, California State University, Fullerton, CA 92834, USA}
\affiliation{$^2$Physikalisch-Technische Bundesanstalt, Bundesallee 100, 38116 Braunschweig, Germany}
\affiliation{$^3$Department of Theoretical Physics, ``Babe\c{s}-Bolyai" University, 40084 Cluj-Napoca, Romania}

\date{\today}
\title{Electronic Green's functions in a T-shaped multi-quantum dot  system}


\begin{abstract}
We developed a set of equations to calculate the electronic Green's functions in a T-shaped multi-quantum dot  system using the equation of motion method. We model the system using a generalized Anderson Hamiltonian which accounts for {\em finite} intradot on-site Coulomb interaction in all component dots as well as for the interdot electron tunneling between adjacent quantum dots. Our results are obtained within and beyond the Hartree-Fock approximation and provide a path to evaluate all the electronic correlations in the multi-quantum dot system in the Coulomb blockade regime. Both approximations provide information on the physical effects related to the finite intradot on-site Coulomb interaction. As a particular example for our generalized results, we considered the simplest T-shaped system consisting of two dots and proved that our approximation introduces important corrections in the detector and side dots Green's functions, and implicitly in the evaluation of the system's transport properties. The multi-quantum dot T-shaped setup may be of interest for the practical realization of qubit states in quantum dots systems.
\end{abstract}
\maketitle

\section{Introduction}

In the last decade, quantum dot (QD) systems provided the perfect environment for the study of complex many-body effects such as the Kondo effect and the Coulomb blockade \cite{tarucha}. For example, electronic transport throughout single QD systems allow the controlled realization of the Kondo regime of the Anderson single impurity problem\cite{cronenwett}.  On the other hand, electronic transport in more complex systems, such as double QD arranged in a series, parallel, or T-shaped configuration \cite{kawakami}, can be explained based on the interplay between Kondo resonances and Fano interference effects \cite{fano}. These sophisticated physical phenomena can be precisely controlled in mesoscopic devices using external parameters such as bias voltages or external electromagnetic fields.
 
Theoretically, a simple model to account for the  system's QD's is the Anderson model \cite{anderson}, each dot being represented by a localized level similar to the impurities in the original problem. Additional terms are introduced to describe the interactions between the system's QD's or the interactions between the QD's and the external leads required by the transport studies. Of main importance is the on-site Coulomb interaction term, which regulate the occupation number in each QD, leading to the direct expression of the Coulomb blockade in transport phenomena\cite{kastner}. Consider the case of a  T-shaped double QD system, with one dot  (detector dot) directly connected to the external leads and the second dot (side dot) coupled to the first one but not to the external leads. Various approximations were applied to estimate the role of the on-site Coulomb interaction in this case.  Wu {\em et al}. \cite{wu} considered an infinite on-site Coulomb interaction in the detector dot  so the double occupancy was forbidden in this dot and neglected it in the side dot. On the other hand, Guclu {\em et al}. \cite{guclu} assumed an infinite on-site Coulomb interaction in the side QD, neglecting it in the detector QD. Tanaka and Kawakami \cite{kawakami} considered the on-site Coulomb interaction to be infinite in both component QD's. Although these configurations are different, both the Kondo effect and the Fano interference effect play an important role in the system's transport properties \cite{,kawakami,wu,guclu}. For each of these studies the main ingredient is the evaluation of the system's QD's Green's functions, as the transport properties depend directly on the detector's dot density of states. 

QD systems were also studied in the nonequilibrium regime \cite{natalya,lavagna,wilczynski}. Nonequilibrium transport studies require special many body techniques based on the Keldysh formalism such as the perturbation theory and the perturbative renormalization group \cite{glazman,parcollet,hooley,wolfle}, slave boson\cite{langreth, meir}, or the equation of motion (EOM)\cite{hewson0}. Additionally, several studies used numerical methods such as the Numerical Renormalization Group\cite{pruschke} or the Quantum Monte Carlo\cite{heary}. Although each of these approaches provided reasonable answers for the transport properties in nonequilibrium mesoscopic systems, all approximations have their limitations. Again, the evaluation of the electronic Green's functions in each of the component QD is the key ingredient of the calculation.

\begin{figure}[tb]
\centering \scalebox{0.5}[0.5]{\includegraphics*{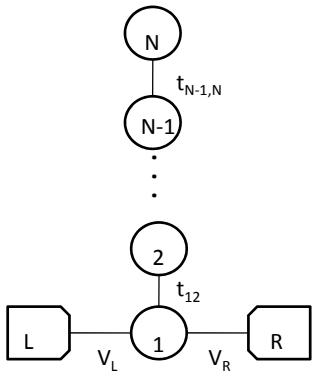}}
\caption{Schematic representation of the T-shape quantum dot system. The detector dot (characteristic energy $\ve_1$) is coupled both to the side quantum dots and the external electrodes $L$ and $R$.}
\label{fig0}
\end{figure}

Here we propose an investigation of the T-shaped  multi-QD's  system (See Fig. \ref{fig0}) based on the EOM method. The system consists on $N$ quantum dots; the detector dot (1) is connected to the external leads $R$ and $L$, while the rest of the dots are connected in a chain arrangement to the detector dot. Our calculation is performed for the most general case when the on-site Coulomb interaction has a {\em finite} value in each of the system's QD's. We consider two different approximations for the evaluation of the system's Green's functions. First, we follow the method introduced by Hewson \cite{hewson} for the study of the Anderson's single impurity model and we obtained results similar to the Hartree-Fock approximation. Second, we consider an approximation beyond the Hartree-Fock approximation which allows us to include higher order effects in the calculation of the system's Green's functions. We derive general recurrence relations which require self-consistent calculations for the evaluation of the electronic Green's function. As an example we will study the double dot system, and prove that the general relations obtained for the general  T-shaped multi-QD's system are  leading to important corrections when one analyzes the system's physical properties. The paper is organized as follows: In Section II we present the general Hamiltonian of the system and the general equations for the system's Green's functions using the EOM method. In Section III we analyze these equations and present their solutions within and beyond the Hartre-Fock approximation. Finally, Section IV presents our conclusions. Details of our calculations are provided in the Appendix.

\section{Model}

The multi-quantum dot T-shaped system is modeled by a generalized Anderson model:
\begin{eqnarray}\label{hamiltonian}
H&=&\sum_{\bk,\sigma;\alpha}\ve_\bk c^\dagger_{\bk\sigma;\alpha}c_{\bk\sigma;\alpha} +\sum_{i=1}^N\ve_i
\sum_\sigma a^\dagger_{i\sigma} a_{i\sigma}\nonumber\\
&+&\sum_{i=1}^N U_i n_{i\sigma}n_{i-\sigma}+\sum_{\bk\sigma;\alpha}V_{\bk 1;\alpha}\left(c^\dagger_{\bk\sigma;\alpha}a_{1\sigma}+
a^\dagger_{1\sigma}c_{\bk\sigma;\alpha}\right)\nonumber\\
&+&\sum_{i=1}^{N-1} t_{i,i+1}\sum_\sigma\left(a^\dagger_{i\sigma}a_{i+1\sigma}+
a^\dagger_{i+1\sigma}a_{i\sigma}\right)\;.
\end{eqnarray}
The first term in the Hamiltonian describes the free electrons in the leads, $c^\dagger_{\bk\sigma;\alpha}$ and $c_{\bk\sigma;\alpha}$ being the fermionic creation and annihilation operators for electrons with momentum $\bk$ and spin $\sigma$ in the lead $\alpha$ ($\alpha\equiv$ left (L), right (R)). The following two terms describe the mesoscopic part of the Hamiltonian and correspond to the electrons localized in the detector ($i=1$) and side ($i=2,\ldots,N$) QD's of the system; here $\ve_i$ is the energy of the localized level and $a^\dagger_{i\sigma}$ and $a_{i\sigma}$ are fermionic creation and annihilation operators for localized electrons with spin $\sigma$ in the $i$'th QD . Additionally, electrons in each component dot are subject to finite on-site Coulomb interaction described by the interaction terms $U_i$; in this interaction term, $n_{i\sigma}=a^\dagger_{i\sigma}a_{i\sigma}$, is the number of particle operator corresponding to the electronic level $\ve_i$ and electron spin $\sigma$. The last two terms in the Hamiltonian describe interactions between the system's electrons. The coupling constant $t_{i,i+1}$ characterizes the electron tunneling between two adjacent dots $i$ and $i+1$. $V_{\bk 1;\alpha}$ characterizes the interaction between the free electrons in the lead $\alpha$ and the localized electrons in the detector dot ($i=1$). For simplicity we will consider the case $V_{\bk 1;L}=V_{\bk 1;R}$ in which the detector couples to the leads only in the symmetric combination $c_{\bk\sigma}=(c_{\bk\sigma;L}+c_{\bk\sigma;R})/\sqrt{2}$ and the dot connects effectively to a single lead, with $V_{\bk 1}=\sqrt{2}V_{\bk 1;L}$.

The system's physical properties can be investigated using the Green's function formalism. In particular, to investigate the transport properties of the system, one will be interested in calculating the detector's dot ($i=1$) localized electrons Green's function. One way to extract the characteristic Green's function for localized electrons in the system's QD's is to use the EOM method.  It is well known that in the case of a general Anderson impurity model, the EOM method leads to an infinite hierarchy of higher-order Green's functions, so in order to obtain the QD's electronic Green's functions one needs to introduce a reliable approximation to truncate this hierarchy. The difficulty is mainly introduced by the interaction terms in the system's Hamiltonian. When the on-site Coulomb interaction term is absent, an exact solution of the problem is possible as it is well known that in this case the set of equations obtained from the EOM method are closed. In the case of two fermionic operators $A$ and $B$ the Fourier transform of the Green's function with respect to the time, $G_{AB}(\omega)=\left<\left<A;B\right>\right>$, is given by the general equation
\be
\omega\left<\left<A;B\right>\right>=\left<\left\{A,B\right\}\right>+\left<\left<\left[A,H\right];B\right>\right>\;,
\ee
where $\left<A\right>$ represents the mean value of the operator $A$, $\left\{A,B\right\}$ the anti-commutator of the operators $A$ and $B$, and $\left[A,B\right]$ their commutator. The last term on the right hand side of the equation is responsible for the generation of the infinite chain of higher order Green's functions.

We start from the evaluation of the electronic Green's functions corresponding to the conduction electrons, $G_{\bk\bk'}^\sigma(\omega)=<<c_{\bk\sigma};c^\dagger_{\bk'\sigma}>>$, and for electrons localized in quantum dot $i$, $G_{ii}^\sigma(\omega)=<<a_{i\sigma};a^\dagger_{i\sigma}>>$. Consider first the free electrons Green's function. Using the general equation, after some simple calculations we find
\begin{equation}\label{free_electrons_Gf}
\left(\omega-\ve_\bk\right) G^\sigma_{\bk\bk'}(\omega)-V_{\bk 1} G^\sigma_{1\bk'}(\omega)=\delta_{\bk\bk'}\;,
\end{equation}
where $\delta_{\bk\bk'}$ is the Kronecker symbol. On the other hand, the localized electrons Green's function will be the solution of the following equation:
\begin{widetext}
\begin{equation}
\label{dot_electrons_GF}
\left(\omega-\ve_i\right) G^\sigma_{ii}(\omega)-U_i\Gamma^\sigma_{ii}(\omega)-\sum_\bk V_{\bk 1} G^\sigma_{\bk 1}(\omega) 
-\Theta(N-1-i) t_{i,i+1}G^\sigma_{i+1,i}(\omega)
-\Theta(i-2) t_{i-1,i} G^\sigma_{i-1,i}(\omega)=1\;,
\end{equation}
\end{widetext}
where $\Theta(x)$ is the standard theta function with $\Theta(x)=1$ for $x\geq 0$ and $\Theta(x)=0$ in rest. One can see from Eqs. (\ref{free_electrons_Gf}) and (\ref{dot_electrons_GF}) that additional correlation functions, i.e., $G^\sigma_{ij}(\omega)$, $G^\sigma_{\bk i}(\omega)$, and $\Gamma^\sigma_{ij}(\omega)$, are introduced by the EOM method and therefore additional equations are required in order to calculate the Green's functions. First, one has to account for correlation functions between electrons in the leads and the QD's, or between electrons from different QD's. Consider first the correlation function between electrons in the leads and electrons in an arbitrary QD, $G^\sigma_{\bk i}(\omega)=<<c_{\bk\sigma};a^\dagger_{i\sigma}>>$. Based on the general equation we find
\begin{equation}\label{Gf_leads_dot}
\left(\omega-\ve_\bk\right)G^\sigma_{\bk i}(\omega)-V_{k1}G^\sigma_{1i}=0\;.
\end{equation}
In a similar way the general equation for the correlation function between electrons in an arbitrary quantum dot and electrons from the leads, $G^\sigma_{i\bk}=<<a_{i\sigma};c^\dagger_{\bk\sigma}>>$, can be obtained as
\begin{widetext}
\begin{equation}\label{Gf_dot_leads}
\left(\omega-\ve_i\right)G^\sigma_{i\bk}(\omega)-U_i\Gamma^\sigma_{i\bk}(\omega)-
\sum_{\bk'}V_{1\bk'}G^\sigma_{\bk'\bk}(\omega)\delta_{i1}
-\Theta(N-1-i) t_{i,i+1} G^\sigma_{i+1,\bk}(\omega)
-\Theta(i-2) t_{i-1,i} G^\sigma_{i-1,\bk}(\omega)=0\;.
\end{equation}
where $\Gamma^\sigma_{i\bk}(\omega)$ is another higher order correlation function.  In addition we have to account for electronic correlations between electrons in two different arbitrary quantum dots, $G^\sigma_{ij}(\omega)=<<a_{i\sigma};a^\dagger_{j\sigma}>>$, a correlation function which can be obtained from the following general equation
\begin{equation}\label{Gf_interdot}
\left(\omega-\ve_i\right)G^\sigma_{ij}(\omega)-U_i\Gamma^\sigma_{ij}(\omega)-\sum_\bk V_{\bk 1} G^\sigma_{\bk j}\delta_{i1}-\Theta(N-1-i) t_{i,i+1} G^\sigma_{i+1,j}(\omega)-\Theta(i-2) t_{i-1,i} G^\sigma_{i-1,j}(\omega)=\delta_{ij}\;.
\end{equation}
Note that for the particular case $i=j$, Eq. (\ref{Gf_interdot}) is identical to Eq. (\ref{dot_electrons_GF}) as expected.

Second, one has to account for additional higher order correlation functions generated by the EOM method. There are two different such functions, namely, $\Gamma^\sigma_{ij}(\omega)=<<n_{i-\sigma}a_{i\sigma};a^\dagger_{j\sigma}>>$ and $\Gamma^\sigma_{i\bk}(\omega)=<<n_{i-\sigma}a_{i\sigma};c^\dagger_{\bk\sigma}>>$. Applying the general formalism of the EOM method one can obtained two general equations for the calculation of these correlation functions
\begin{eqnarray}\label{gamma_ij}
\left(\omega-\ve_i-U_i\right)\Gamma^\sigma_{ij}(\omega)-\sum_\bk V_{\bk1}\delta_{i1}\left[\left<\left<n_{1-\sigma}c_{\bk\sigma};a^\dagger_{j\sigma}\right>\right>+
\left<\left<a^\dagger_{1-\sigma}c_{\bk-\sigma}a_{1\sigma};a^\dagger_{j\sigma}\right>\right>-
\left<\left<c^\dagger_{\bk-\sigma}a_{1-\sigma}a_{1\sigma};a^\dagger_{j\sigma}\right>\right>\right]&&\nonumber\\
-\Theta(N-1-i) t_{i,i+1}\left[\left<\left<n_{i-\sigma}a_{i+1\sigma};a^\dagger_{j\sigma}\right>\right>
+\left<\left<a^\dagger_{i-\sigma}a_{i+1-\sigma}a_{i\sigma};a^\dagger_{j\sigma}\right>\right>-
\left<\left<a^\dagger_{i+1-\sigma}a_{i-\sigma}a_{i\sigma};a^\dagger_{j\sigma}\right>\right>\right]&&\nonumber\\
-\Theta(i-2) t_{i-1,i}\left[\left<\left<n_{i-\sigma}a_{i-1\sigma};a^\dagger_{j\sigma}\right>\right>+
\left<\left<a^\dagger_{i-\sigma}a_{i-1-\sigma}a_{i\sigma};a^\dagger_{j\sigma}\right>\right>-
\left<\left<a^\dagger_{i-1-\sigma}a_{i-\sigma}a_{i\sigma};a^\dagger_{j\sigma}\right>\right>\right]&=&
\left<n_{i-\sigma}\right>\delta_{ij}\;\nonumber\\
\end{eqnarray}
and
\begin{eqnarray}\label{gamma_ik}
\left(\omega-\ve_i-U_i\right)\Gamma^\sigma_{i\bk}(\omega)-\sum_\bk V_{\bk1}\delta_{i1}\left[\left<\left<n_{1-\sigma}c_{\bk\sigma};c^\dagger_{\bk\sigma}\right>\right>+
\left<\left<a^\dagger_{1-\sigma}c_{\bk-\sigma}a_{1\sigma};c^\dagger_{\bk\sigma}\right>\right>-
\left<\left<c^\dagger_{\bk-\sigma}a_{1-\sigma}a_{1\sigma};c^\dagger_{\bk\sigma}\right>\right>\right]&&\nonumber\\
-\Theta(N-1-i) t_{i,i+1}\left[\left<\left<n_{i-\sigma}a_{i+1\sigma};c^\dagger_{\bk\sigma}\right>\right>
+\left<\left<a^\dagger_{i-\sigma}a_{i+1-\sigma}a_{i\sigma};c^\dagger_{\bk\sigma}\right>\right>-
\left<\left<a^\dagger_{i+1-\sigma}a_{i-\sigma}a_{i\sigma};c^\dagger_{\bk\sigma}\right>\right>\right]&&\nonumber\\
-\Theta(i-2) t_{i-1,i}\left[\left<\left<n_{i-\sigma}a_{i-1\sigma};c^\dagger_{\bk\sigma}\right>\right>+
\left<\left<a^\dagger_{i-\sigma}a_{i-1-\sigma}a_{i\sigma};c^\dagger_{\bk\sigma}\right>\right>-
\left<\left<a^\dagger_{i-1-\sigma}a_{i-\sigma}a_{i\sigma};c^\dagger_{\bk\sigma}\right>\right>\right]&=& 0\;.
\end{eqnarray}
\end{widetext}
As expected, the EOM method introduces additional higher order correlation functions. Basically, one can continue to apply the same method for the calculation of all higher order correlation functions, with the hope that the resulting chain of equations will close at a certain level and therefore a solution for the system's Green's functions can be obtained. In practice it is well known that such a procedure will lead to an infinite number of self consistent equations, and to obtain a set of solutions for the system's Green's function appropriate approximations has to be done.

\section{Approximation methods}
Several different approximations were used in connection with the EOM method. From these. probably the simplest one is the Hartree-Fock approximation. If used for the Anderson impurity model, the Hartree-Fock approximation simplifies the on-site Coulomb interaction term by replacing the number of particles operators with their average values, $U_i<n_{i\sigma}><n_{i-\sigma}>$. A similar approximation was used by Hewson to describe the localized magnetic states in metals\cite{hewson}. Lacroix used a higher order approximation to explain the Kondo effect in the impurity Anderson model \cite{lacroix}.
\subsection{Hartree-Fock approximation}

One of the most common approximation used in many-body theory of fermionic systems is the so called Hartree-Fock approximation. This approximation was used successfully by Hewson \cite{hewson} to investigate the single impurity Anderson model. Let us analyze the higher order correlation functions generated in the right-hand-side of Eqs. (\ref{gamma_ij}) and (\ref{gamma_ik}), as several of these terms are neglected in the Hartree-Fock approximation. The first category of terms we encounter are the so called normal scattering correlation functions. In the case of the $\Gamma^\sigma_{ij}(\omega)$ correlation function these terms are $<<n_{i-\sigma}c_{\bk\sigma};a^\dagger_{j\sigma}>>\delta_{i1}$ and $<<n_{i-\sigma}a_{l\sigma};a^\dagger_{j\sigma}>>$ ($l$=$i-1$ or $l$=$i+1$). The first term describes the scattering of the conduction electrons by the localized electrons in the detector dot with the generation of a localized electron in the QD $i=2$. The second term describes the scattering  of a localized electron in the QD $i-1$ ($i+1$) by the electrons in the adjacent dot $i$, with a generation of an electron in the QD $i+1$ ($i-1$). In the Hartree-Fock approximation these correlation functions are replaced by a product of the average occupation number  and an additional correlation function $<n_{i-\sigma}><<c_{\bk\sigma};a^\dagger_{j\sigma}>>$ and $<n_{i-\sigma}><<a_{l\sigma};a^\dagger_{j\sigma}>>$, respectively. The remaining terms are neglected in the Hartree-Fock approximation. A similar approximation is used for $\Gamma^\sigma_{i\bk}(\omega)$. For this correlation function the normal scattering terms  $<<n_{i-\sigma}c_{\bk\sigma};c^\dagger_{\bk\sigma}>>$ and $<<n_{i-\sigma}a_{l\sigma};c^\dagger_{\bk\sigma}>>$ ($l$=$i-1$ or $l$=$i+1$) are replaced by $<n_{i-\sigma}><<c_{\bk\sigma};c^\dagger_{\bk\sigma}>>$ and $<n_{i-\sigma}><<a_{l\sigma};c^\dagger_{\bk\sigma}>>$, respectively. Again, one neglects all the remaining terms in the Hartree-Fock approximation. Accordingly, the two higher order correlation functions can be expressed in terms of lower order correlation functions
\begin{widetext}
\begin{equation}\label{gamma_ij_HF}
\Gamma^\sigma_{ij}(\omega)=\left<n_{i-\sigma}\right>\f{\delta_{ij}+\sum_\bk V_{k1}G^\sigma_{\bk j}(\omega)\delta_{i1}+\Theta(N-1-i) t_{i,i+1}G^\sigma_{i+1,j}(\omega)+\Theta(i-2)G^\sigma_{i-1,j}(\omega)}{\omega-\ve_i-U_i}
\end{equation}
and
\begin{equation}\label{gamma_ik_HF}
\Gamma^\sigma_{i\bk}(\omega)=\left<n_{i-\sigma}\right>\f{\sum_\bk V_{k1}G^\sigma_{\bk\bk}(\omega)\delta_{i1}+\Theta(N-1-i) t_{i,i+1}G^\sigma_{i+1,\bk}(\omega)+\Theta(i-2)G^\sigma_{i-1,\bk}(\omega)}{\omega-\ve_i-U_i}\;.
\end{equation}
We can use these approximated higher order correlation functions along with Eqs. (\ref{free_electrons_Gf}) -- (\ref{Gf_interdot}) to calculate the Green's functions for the localized electrons in the system's QD's. After some simple algebra, a general recurrence relation can be obtained in the form
\begin{equation}\label{recurrence_Gij}
A_i(\omega) G^\sigma_{ij}(\omega)-\delta_{i1} \Sigma_\bk^0(\omega)G^\sigma_{1j}(\omega)-
\Theta(N-1-i) t_{i,i+1} G^\sigma_{i+1,j}(\omega)-\Theta(i-2) t_{i-1,i} G^\sigma_{i-1,j}=\delta_{ij}\;,
\end{equation}
\end{widetext}
with
\begin{equation}\label{defSigma0}
\Sigma_\bk^0(\omega)=\sum_\bk\f{\left|V_{1\bk}\right|^2}{\omega-\ve_\bk}
\end{equation}
and
\begin{equation}\label{defA}
A_i(\omega)=\f{\left(\omega-\ve_i\right)(\omega-\ve_i-U_i)}{\omega-\ve_i-U_i\left(1-\left<n_{i-\sigma}\right>\right)}\;.
\end{equation}
The above equation gives us a direct relation between the electronic Green's functions $G^\sigma_{ij}(\omega)$, $G^\sigma_{i+1,j}(\omega)$, and $G^\sigma_{i-1,j}(\omega)$.

Let us first consider the Green's function for the detector dot, $G^\sigma_{11}(\omega)$. First we set $i=1$ and $j=1$ in Eq. (\ref{recurrence_Gij})
\begin{equation}\label{G11_HF}
\left[A_1(\omega)-\Sigma_\bk^0(\omega) \right]G^\sigma_{11}(\omega)-t_{12} G^\sigma_{21}=1\;.
\end{equation}
To find the general form of the interdot Green's function $G^\sigma_{21}$ we can use the complete chain of equations generated for arbitrary $i$ ($i=2,\ldots,N$) and $j=1$. Based on this procedure one finds
\begin{eqnarray}
\left[G^\sigma_{11}(\omega)\right]^{-1}&=&A_1(\omega)-\Sigma_\bk^0(\omega)\nonumber\\
&&-\f{t^2_{12}}{A_2(\omega)-
\f{t^2_{23}}{\hspace{1cm}\f{\ddots}{A_{N-1}(\omega)-\f{t^2_{N-1,N}}{A_N(\omega)}}}}\;,\;\;\;\;
\end{eqnarray}
an equation which is in complete agreement with previous results obtained in the Hartree-Fock approximation for systems with one or two dots \cite{hewson,tifrea}. In a similar way, for the final dot of the system ($i=N$) we find
\begin{eqnarray}
&&\left[G^\sigma_{NN}(\omega)\right]^{-1}\nonumber\\
&&=A_N(\omega)-\f{t^2_{N-1,N}}{A_{N-1}(\omega)-
\f{t^2_{N-2,N-1}}{\hspace{1cm}\f{\ddots}{A_{2}(\omega)-\f{t^2_{12}}{A_1(\omega)-\Sigma_\bk^0(\omega)}}}}\;.\;\;\;\;
\end{eqnarray}
Finally, let us consider the general case of an intermediate QD, $1<j<N$. Following a similar procedure, in the Hartree-Fock approximation we find
\begin{eqnarray}\label{Gf_finalHF}
\left[G^\sigma_{jj}(\omega)\right]^{-1}&=&A_j(\omega)\nonumber\\
&&-\f{t^2_{j,j+1}}{A_{j+1}(\omega)-
\f{t^2_{j+1,j+2}}{\hspace{1cm}\f{\ddots}{A_{N-1}(\omega)-\f{t^2_{N-1,N}}{A_N(\omega)}}}}\nonumber\\
&&-\f{t^2_{j-1,j}}{A_{j-1}(\omega)-
\f{t^2_{j-2,j-1}}{\hspace{1cm}\f{\ddots}{A_{2}(\omega)-\f{t^2_{12}}{A_1(\omega)-\Sigma_\bk^0(\omega)}}}}\;.\;\;\;\;\;\;\;\;
\end{eqnarray}
Although the general equation for the $j$'th QD Green's function looks relatively simple, in fact we have to consider a set of self-consistent equations as the general factor $A_i(\omega)$ depends on the average occupancy of the $i$'th QD, $\left<n_{i-\sigma}\right>$ (see Eq. (\ref{defA})), a value which can be evaluated using the general relation
\begin{equation}\label{occupancy}
\left<n_{i-\sigma}\right>=-\f{1}{\pi}\int_{-\infty}^{\infty} f(\omega)\textrm{Im}G_{ii}^{-\sigma}(\omega) d\omega\;,
\end{equation}
$f(\omega)$ being the standard Fermi-Dirac function. Moreover, one can see that the two spin channels, $\sigma$ and $-\sigma$, are coupled in Eq. (\ref{Gf_finalHF}). Such a property makes the self-consistent set of equations difficult to solve exactly. The easiest way to solve this problem is to start from the non-interacting case ($U_i=0$), situation in which the set of equations fully decouples, and an exact solution is possible. For example, such solutions were obtained for the case of double or triple quantum dot systems \cite{tifrea,japonezi}. The same results can be obtained using our general formalism. The quantum dots Green's functions for the non-interacting case can be used as a starting point in a numerical self-consistent calculation for the interacting case.

\subsection{Beyond Hartree-Fock approximation}
The Hartree-Fock approximation lacks to explain many physical properties of the system. For example, in the case of the Anderson impurity model, Lacroix proved that the understanding of the Kondo effect requires a higher order approximation \cite{lacroix}. To obtain results beyond the Hartree-Fock approximation, the terms in the right hand sides of Eqs. (\ref{gamma_ij}) and (\ref{gamma_ik}) have to be evaluated using the EOM method. The calculations are lengthly, but not extremely complicated, the only downside being that the EOM method will introduce even higher order correlation functions which thereafter have to be approximated. The standard approach is to neglect all higher correlation terms, i.e., terms which introduce correlations between more that four fermionic operators.

Let us consider for example the equation for the correlation function $\Gamma^\sigma_{ij}(\omega)$. Eq. (\ref{gamma_ij}) introduces an additional nine higher order correlation functions. We will have to use the EOM method to evaluate and thereafter to approximate these functions. The general equations for these correlation functions are presented in the Appendix \ref{AppendixA}. Accordingly, one finds:

\begin{widetext}
\begin{eqnarray}
\Gamma_{ij}^\sigma(\omega)&=&\left<n_{i-\sigma}\right>
\f{\delta_{ij}+\delta_{i2}\Theta(i-2)t_{i-1,i}\f{\sum_{\bk'}V_{\bk' 1}G^\sigma_{\bk' j} (\omega)}{\omega-\ve_{i-1}}+\Theta(N-1-i)\f{t_{i,i+1}}{\omega-\ve_{i+1}}\delta_{i+1,j}+\Theta(i-2)\f{t_{i-1,i}}{\omega-\ve_{i-1}}\delta_{i-1,j}}
{\omega-\ve_i-U_i-\delta_{i1}\left[2\Sigma^0_\bk (\omega)+\Sigma_\bk^1(\omega)\right]-\Theta(N-1-i)t^2_{i,i+1}M_{i,i+1}(\omega)-\Theta(i-2)t^2_{i-1,i}M_{i,i-1}(\omega)}\nonumber\\
&&+\left<n_{i-\sigma}\right>
\f{\Theta(N-2-i)\f{t_{i+1,i+2}}{\omega-\ve_{i+1}}G^\sigma_{i+2,j}(\omega)+\Theta(i-3)\f{t_{i-2,i-1}}{\omega-\ve_{i-1}}G^\sigma_{i-2,j}(\omega)}
{\omega-\ve_i-U_i-\delta_{i1}\left[2\Sigma^0_\bk (\omega)+\Sigma_\bk^1(\omega)\right]-\Theta(N-1-i)t^2_{i,i+1}M_{i,i+1}(\omega)-\Theta(i-2)t^2_{i-1,i}M_{i,i-1}(\omega)}\nonumber\\
&&-G_{ij}^\sigma(\omega)\f{\delta_{i1}\Sigma_\bk^2(\omega)+\Theta(N-1-i)t^2_{i,i+1}\left<n_{i+1-\sigma}\right>P_{i,i+1}(\omega)-\Theta(i-2)t^2_{i-1,i}\left<n_{i-1-\sigma}\right>P_{i,i-1}(\omega)}
{\omega-\ve_i-U_i-\delta_{i1}\left[2\Sigma^0_\bk (\omega)+\Sigma_\bk^1(\omega)\right]-\Theta(N-1-i)t^2_{i,i+1}M_{i,i+1}(\omega)-\Theta(i-2)t^2_{i-1,i}M_{i,i-1}(\omega)}\nonumber\\
&&+\f{\delta_{i1}A^\sigma_{ij}(\omega)+\Theta(N-1-i)t_{i,i+1}B_{ij}^\sigma(\omega)+
\Theta(i-2)t_{i-1,i}C^\sigma_{ij}(\omega)}
{\omega-\ve_i-U_i-\delta_{i1}\left[2\Sigma^0_\bk (\omega)+\Sigma_\bk^1(\omega)\right]-\Theta(N-1-i)t^2_{i,i+1}M_{i,i+1}(\omega)-\Theta(i-2)t^2_{i-1,i}M_{i,i-1}(\omega)}\;.
\end{eqnarray}
A similar expression can be obtained also for $\Gamma^\sigma_{i\bk}(\omega)$. The above equation can be combined with Eqs. (\ref{Gf_leads_dot}) and (\ref{Gf_interdot}) to obtain a generalized recurrence relation for the electronic inter-dot Green's function:
\begin{eqnarray}\label{Gf_BHF_interdot}
&&\left(\omega-\ve_i+U_i\f{\delta_{i1}\Sigma_\bk^2(\omega)+\Theta(N-1-i)t^2_{i,i+1}\left<n_{i+1-\sigma}\right>P_{i,i+1}(\omega)-\Theta(i-2)t^2_{i-1,i}\left<n_{i-1-\sigma}\right>P_{i,i-1}(\omega)}{\omega-\ve_i-U_i-\delta_{i1}\left[2\Sigma^0_\bk (\omega)+\Sigma_\bk^1(\omega)\right]-\Theta(N-1-i)t^2_{i,i+1}M_{i,i+1}(\omega)-\Theta(i-2)t^2_{i-1,i}M_{i,i-1}(\omega)}\right) G^\sigma_{ij}(\omega)\nonumber\\
&&-\left(\delta_{i1}\Sigma_\bk^0(\omega) -
\f{\delta_{i2}\Sigma_\bk^0(\omega)\Theta(i-2)\f{t_{i-1,i}}{\omega-\ve_{i-1}}U_i\left<n_{i-\sigma}\right>}
{\omega-\ve_i-U_i-\delta_{i1}\left[2\Sigma^0_\bk (\omega)+\Sigma_\bk^1(\omega)\right]-\Theta(N-1-i)t^2_{i,i+1}M_{i,i+1}(\omega)-\Theta(i-2)t^2_{i-1,i}M_{i,i-1}(\omega)}\right)G^\sigma_{1j}(\omega)\nonumber\\
&&-\Theta(N-1-i) t_{i,i+1} G^\sigma_{i+1,j}(\omega)-\Theta(i-2) t_{i-1,i} G^\sigma_{i-1,j}(\omega)\nonumber\\
&&-\f{\Theta(N-2-i)\f{t_{i+1,i+2}}{\omega-\ve_{i+1}}U_i\left<n_{i-\sigma}\right>}{\omega-\ve_i-U_i-\delta_{i1}\left[2\Sigma^0_\bk (\omega)+\Sigma_\bk^1(\omega)\right]-\Theta(N-1-i)t^2_{i,i+1}M_{i,i+1}(\omega)-\Theta(i-2)t^2_{i-1,i}M_{i,i-1}(\omega)} G^\sigma_{i+2,j}(\omega)\nonumber\\
&&-\f{\Theta(i-3)\f{t_{i-2,i-1}}{\omega-\ve_i}U_i\left<n_{i-\sigma}\right>}
{\omega-\ve_i-U_i-\delta_{i1}\left[2\Sigma^0_\bk (\omega)+\Sigma_\bk^1(\omega)\right]-\Theta(N-1-i)t^2_{i,i+1}M_{i,i+1}(\omega)-\Theta(i-2)t^2_{i-1,i}M_{i,i-1}(\omega)}G^\sigma_{i-2,j}(\omega)\nonumber\\
&&-U_i\f{\delta_{i1}A^\sigma_{ij}(\omega)+\Theta(N-1-i)t_{i,i+1}B_{ij}^\sigma(\omega)+
\Theta(i-2)t_{i-1,i}C^\sigma_{ij}(\omega)}
{\omega-\ve_i-U_i-\delta_{i1}\left[2\Sigma^0_\bk (\omega)+\Sigma_\bk^1(\omega)\right]-\Theta(N-1-i)t^2_{i,i+1}M_{i,i+1}(\omega)-\Theta(i-2)t^2_{i-1,i}M_{i,i-1}(\omega)}\nonumber\\
&&=\delta_{ij}\left[1+\f{U_i\left<n_{i-\sigma}\right>}{\omega-\ve_i-U_i-\delta_{i1}\left[2\Sigma^0_\bk (\omega)+\Sigma_\bk^1(\omega)\right]-\Theta(N-1-i)t^2_{i,i+1}M_{i,i+1}(\omega)-\Theta(i-2)t^2_{i-1,i}M_{i,i-1}(\omega)}\right]\nonumber\\
&&+U_i\left<n_{i-\sigma}\right>\f{\Theta(N-1-i)\f{t_{i,i+1}}{\omega-\ve_{i+1}}\delta_{i+1,j}+\Theta(i-2)\f{t_{i-1,i}}{\omega-\ve_{i-1}}\delta_{i-1,j}}
{\omega-\ve_i-U_i-\delta_{i1}\left[2\Sigma^0_\bk (\omega)+\Sigma_\bk^1(\omega)\right]-\Theta(N-1-i)t^2_{i,i+1}M_{i,i+1}(\omega)-\Theta(i-2)t^2_{i-1,i}M_{i,i-1}(\omega)}\;.
\end{eqnarray}
\end{widetext}
The above equation in its most general form it is a self-consistent integral equation. Different from the Hartree-Fock approximation case, here, the inter-dot electronic Green's function $G_{ij}^\sigma(\omega)$ depends on the inter-dot electronic Green's functions for the near neighbors $G_{i\pm1,j}^\sigma(\omega)$ and the  inter-dot electronic Green's functions for the next-near neighbors $G^\sigma_{i\pm 2,j}(\omega)$. Similar to the Hartree-Fock approximation, the inter-dot electronic Green's function depends on the average electronic occupancy  $\left<n_{i-\sigma}\right>$ in dot $i$, which can be calculated using the general relation (\ref{occupancy}). Eq. (\ref{Gf_BHF_interdot}) can be used to obtain the intra-dot electronic Green's function $G_{ii}^\sigma(\omega)$ by simply setting $j=i$. Additionally, anomalous terms ($A_{ij}^\sigma(\omega)$, $B_{ij}^\sigma(\omega)$,  and $C_{ij}^\sigma(\omega)$) are present in the general equation for the electronic Green's function Eq. (\ref{Gf_BHF_interdot}). These terms will introduce new average values of the type $\left<a^\dagger_{i-\sigma} c_{\bk -\sigma}\right>$, $\left<c^\dagger_{\bk-\sigma} a_{i-\sigma}\right>$, and $\left<a^\dagger_{i-\sigma} a_{j-\sigma}\right>$, which have to be calculated self-consistently. For example, one can calculate $\left<a^\dagger_{i-\sigma} c_{\bk-\sigma}\right>$ as
\begin{equation}\label{AnAve1}
\left<a^\dagger_{i-\sigma} c_{\bk-\sigma}\right>=-\f{1}{\pi}\int_{-\infty}^\infty f(\omega) \textrm{Im}G_{\bk i}^{-\sigma}(\omega)\;,
\end{equation}
or $\left<a^\dagger_{i-\sigma} a_{j-\sigma}\right>$ as
\begin{equation}\label{AnAve2}
\left<a^\dagger_{i-\sigma} a_{j-\sigma}\right>=-\f{1}{\pi}\int_{-\infty}^\infty f(\omega) \textrm{Im}G_{ij}^{-\sigma}(\omega) d\omega\;.
\end{equation}
Although neglected in some quantum dot systems studies \cite{natalya}, such anomalous averages have proven to be of main importance in the investigation of  the system's physical properties \cite{lacroix,yunong}. 

\section{T-shaped double quantum dot  system in the Coulomb blockade regime}

The simplest T-shape quantum dot system described by the model is the double quantum dot system. Simply, if we set $t_{23}=0$, the only remaining components of our system are the quantum dot connected to the external leads (detector dot) and a second quantum dot connected to the first one, but not to the external leads (side dot).  In the following we will investigate the electronic density of states in the detector quantum dot ($\rho^\sigma_1(\omega)=-\textrm{Im} G^\sigma_{11}(\omega)/\pi$) and its dependence on the on-site Coulomb interaction terms $U_1$ and $U_2$ using various approximations. First, we will point to various levels of approximation for this system and thereafter we will present numerical results for the occupation number of the system. 

The simplest possible situation is for $U_1=U_2=0$, case in which the density of states for the detector dot can be calculated analytically as:
\begin{equation}
\rho^\sigma_1(\omega)=\f{1}{\pi}\f{\Delta}{\left(\omega-\ve_1-\f{t_{12}^2}{\omega-\ve_2}\right)^2+\Delta^2}\;,
\end{equation}
where $\Delta$ is the imaginary part of  $\Sigma_\bk^0(\omega+i\eta)$ ($\eta\rightarrow 0$),i.e.,
\begin{displaymath}
\Delta=\pi\sum_\bk |V_{\bk 1}|^2\delta(\omega-\ve_\bk)\;,
\end{displaymath}
and $\delta(x)$ is the delta Dirac function. In this situation, the electronic density of states will present a double peak structure, the structure and the position of these peaks being controlled by the system's parameters \cite{tifrea}. Let us consider the case of finite on-site Coulomb interaction in both the detector ($U_1\neq 0$) and side ($U_2\neq 0$) dots.

\subsection{Hartree-Fock Approximation}

The general theory presented in the previous section allow us to estimate the detector's Green's function within the Hartree-Fock approximation:

\begin{eqnarray}\label{DD_G11}
&&G_{11}^\sigma(\omega)=\nonumber\\
&&\f{1}{\f{\left(\omega-\ve_1\right)\left(\omega-\ve_1-U_1\right)}{\omega-\ve_1-U_1\left(1-\left<n_{1-\sigma}\right>\right)}-t_{12}^2\f{\omega-\ve_1-U_2\left(1-\left<n_{2-\sigma}\right>\right)}{\left(\omega-\ve_2\right)\left(\omega-\ve_2-U_2\right)}-\Sigma_\bk^0(\omega)}\;\;\;.\nonumber\\
\end{eqnarray}
Although the equation for the detector's dot Green's function is analytic, before one can estimate the detector dot density of states $\rho_1^\sigma(\omega)$, one has to self-consistently evaluate the average occupation number for both the detector and side dots. In the case of the side dot, the required Green's function is given as

\begin{eqnarray}\label{DD_G22}
&&G_{22}^\sigma(\omega)=\nonumber\\
&&\f{1}{\f{\left(\omega-\ve_1\right)\left(\omega-\ve_2-U_2\right)}{\omega-\ve_2-U_2\left(1-\left<n_{2-\sigma}\right>\right)}-\f{t_{12}^2}{\f{\left(\omega-\ve_1\right)\left(\omega-\ve_1-U_1\right)}{\omega-\ve_1-U_1\left(1-\left<n_{1-\sigma}\right>\right)}-\Sigma_\bk^0(\omega)}}\;\;.\nonumber\\
\end{eqnarray}
Both occupation numbers can be calculated using Eq. (\ref{occupancy}) by means of an iterative numerical evaluation. The resulting density of states presents additional peaks as a result of non-zero on-site Coulomb interaction terms.

\subsection{Beyond Hartree-Fock Approximation}

In general, the Hartree-Fock approximation can give some insight on the importance of the on-site Coulomb interaction terms $U_1$ and $U_2$, however, it is necessary to go beyond this approximation to recover important physical properties of the system. For example, Lacroix \cite{lacroix} proved that a calculation of the Kondo temperature using the single impurity Anderson model in the $U_1\rightarrow\infty$ limit requires terms beyond the Hartree-Fock approximation. We expect that such terms will be relevant also to the finite $U_1$ and $U_2$ case. To calculate the detector and side dots Green's functions we will need to use a set of coupled equations:

\begin{widetext}
\begin{eqnarray}\label{G11_BHF}
&&\left(\omega-\ve_1-\Sigma_\bk^0(\omega)+U_1\f{\Sigma_\bk^2(\omega)+t^2_{12}\left<n_{2-\sigma}\right>P_{12}(\omega)}{\omega-\ve_1-U_1-\left[2\Sigma_\bk^0(\omega)+\Sigma_\bk^1(\omega)\right]-t^2_{12}M_{12}(\omega)}\right) G_{11}^\sigma(\omega)\nonumber\\
&&-t_{12} G^\sigma_{21}(\omega)-U_1\f{A^\sigma_{11}(\omega)+t_{12} B^\sigma_{11}(\omega)}{\omega-\ve_1-U_1-\left[2\Sigma_\bk^0(\omega)+\Sigma_\bk^1(\omega)\right]-t^2_{12}M_{12}(\omega)}\nonumber\\
&&=1+\f{U_1\left<n_{1-\sigma}\right>}{\omega-\ve_1-U_1-\left[2\Sigma_\bk^0(\omega)+\Sigma_\bk^1(\omega)\right]-t^2_{12}M_{12}(\omega)}
\end{eqnarray}
and
\begin{eqnarray}\label{G22_BHF}
&&\left(\omega-\ve_2-U_2\f{t^2_{12}\left<n_{1-\sigma}\right>P_{21}(\omega)}{\omega-\ve_2-U_2-t^2_{12}M_{21}(\omega}\right)G^\sigma_{22}(\omega)-t_{12} G^\sigma_{12}(\omega)\nonumber\\
&&+\f{t_{12}}{\omega-\ve_1}
\f{\Sigma^0_\bk(\omega)U_2\left<n_{2-\sigma}\right>}{\omega-\ve_2-U_2-t^2_{12} M_{21}(\omega)}
-U_2\f{t_{12}C_{22}^\sigma(\omega)}{\omega-\ve_2-U_2-t^2_{12}M_{21}(\omega)}\nonumber\\
&&=1+\f{U_2\left<n_{2-\sigma}\right>}{\omega-\ve_2-U_2-t^2_{12}M_{21}(\omega)}\;.
\end{eqnarray}
\end{widetext}

Clearly, the above equations are self-consistent due to the presence of terms involving the occupation numbers in the detector and side dots. Moreover, the presence of additional correlation functions, $G_{12}^\sigma(\omega)$ and $G_{21}^\sigma(\omega)$, makes the calculation even more complicated. Additional equations are required to replace $G_{12}^\sigma(\omega)$ and $G_{21}^\sigma(\omega)$ in terms of $G_{11}^\sigma(\omega)$ and $G_{22}^\sigma(\omega)$. After some algebraic manipulations, both detector and side dots Green's functions can be expressed only in terms of the average occupation numbers $\left<n_{1\sigma}\right>$ and $\left<n_{2\sigma}\right>$ and a series of anomalous averages of the type $\left<a^\dagger_{1-\sigma}c_{\bk-\sigma} \right>$, $\left<c^\dagger_{\bk'-\sigma} c_{\bk-\sigma}\right>$, $\left<a^\dagger_{2-\sigma}c_{\bk-\sigma} \right>$, and $\left<a^\dagger_{2-\sigma}a_{1-\sigma} \right>$, along with their hermitian conjugates. The presence of these terms makes the problem self-consistent as such averages have to be calculated using their corresponding Green's functions using Eqs. (\ref{occupancy}), (\ref{AnAve1}), and (\ref{AnAve2}).

Two levels of approximation can be considered in this limit. First, one can consider all the anomalous averages to be zero. A similar procedure was used by Zimbovskaya \cite{natalya} in connection with the Coulomb blockade regime in single quantum dot systems. On the other hand, Lacroix proved that the Kondo regime ($U_1\rightarrow \infty$) of the single impurity Anderson model can be explain only if the anomalous averages are taken into account  \cite{lacroix}. Our calculation can be reduced to the single impurity Anderson model if we consider the limit $t_{12}\rightarrow 0$, i.e., we decouple the side dot of the T-shape system. The obtained Green's function for the detector dot matches the previous findings of both Zimbovskaya and Lacroix, assuming the right levels of approximation.

In the case of the double dot T-shape system, for finite onsite Coulomb interaction in both the detector and side dots, one has to use numerical calculations to evaluate the system's properties. We choose to calculate the system's average occupation number, $n_{t,\sigma}$=$\left<n_{1\sigma}\right>$+$\left<n_{2\sigma}\right>$. Our main goal is to prove the differences between the two levels of approximation respect to the anomalous averages mentioned above. First, we followed Zimbovskaya and we disregarded these averages (BHF1). Second, we followed Lacroix and we took into account these averages (BHF2), however, in this limit we considered a more general situation with finite on-site Coulomb interaction in both component dots (not only the particular situation with infinite on-site Coulomb intercation).

\begin{figure}[b]
\includegraphics[trim=005 005 005 005,clip,width=\columnwidth]{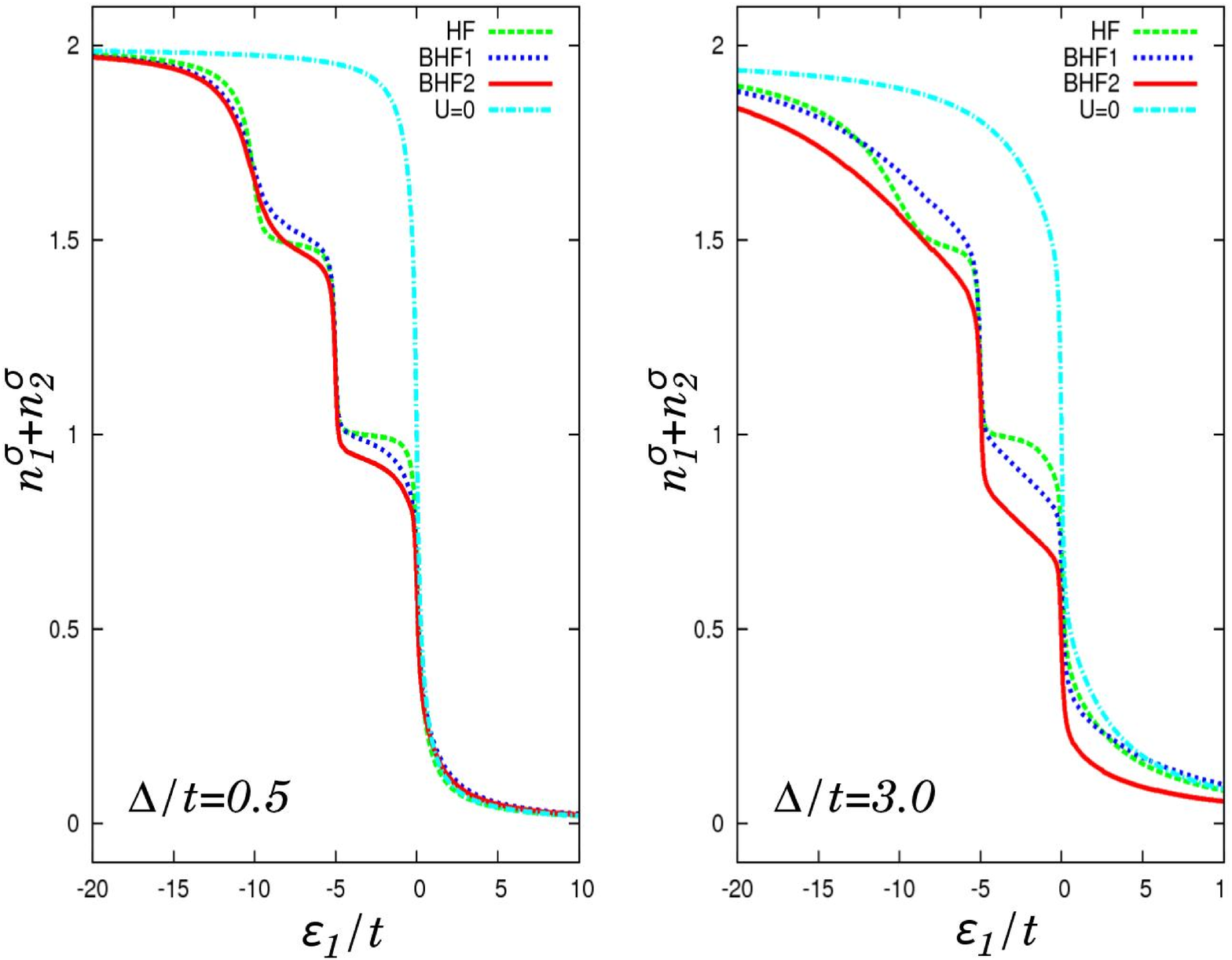}
\caption{The system's average total occupation number (per spin orientation) as function of the localized energy level  in the detector dot in various approximations: Hartree-Fock approximation (dashed line), beyond Hartree-Fock approximation without (dotted line) and with (dot-dashed line) considering all anomalous averages. The considered parameters are
$U_1/t=10$, $U_2/t=5$, $E_F/t=0$, $T/t=10^{-4}$,  $\Delta/t=0.5$ (left), and $\Delta/t=3.0$ (right). We also considered  $\ve_1/t=\ve_2/t$. For comparison, we present results in the simple case without on-site Coulomb interaction $U_1=U_2=0$ (dot-dashed line).}
\label{fig2}
\end{figure}

Figure \ref{fig2} showcases a numerical estimation of the electron occupation number per spin orientation of the double dot T-shape system. We considered two different situations, $\Delta/t=0.5$ and $\Delta/t=3$, corresponding to the case of a slow (weak coupling) and fast (strong coupling) detector configurations. The plot presents the Hartree-Fock  approximation results (dashed line)  and the results obtained using terms beyond the  Hartree-Fock approximation without - similar to the approximation in Ref. \onlinecite{natalya} -  (dotted line) and with (dot-dashed line) considering all anomalous averages in the detector and side dots  Green's function. For comparison, we plotted also the simple case without on-site Coulomb interaction, $U_1=U_2=0$ (dot-dashed line), situation in which the Coulomb blockade regime is not possible. Clearly, the inclusion of the anomalous terms in the calculation resulted in significant changes in the system's total average occupation number, especially in the Coulomb blockade regime.  When the energy of the localized level in the detector dot is well bellow the Fermi energy of the free electrons on the leads ($E_F=0$), one electron per spin orientation will be accommodated on each localized level of the system. As the energy increases, the effects of the on-site Coulomb interaction are becoming important, so only one electron can occupy the localized levels. Once the energy of the localized level raises above the Fermi energy, the occupation of the localized levels approaches zero as electrons will flow directly from one lead to the other. In principle, the relative value of the localized level characteristic energy can be controlled using an external gate.

\section{Conclusions}

In conclusion, we presented a detailed analysis of a T-shaped multi-quantum dot  systems using the equation of motion method. All our calculations are done in the case of a {\em finite} on-site Coulomb interaction in each of the component dots. The results are obtained within various approximations and for each case we provided a set of self-consistent equations which allow the calculation of the electronic Green's function in the system's component dots. The simplest approximation is similar to the one introduced for a single quantum dot system by Hewson \cite{hewson} and is equivalent to the Hartree-Fock approximation. In this case, the general recurrence equation which relates the electronic Green's function between adjacent quantum dots in the system can be solved and we obtained a set of self-consistent equations for the electronic Green's functions. The self-consistency in the case of the Hartree-Fock approximation is introduced throughout the electronic occupation number in each of the component dots. When an approximation which includes terms beyond the Hartree-Fock approximation is considered, the situation is more complex and the analysis of the results is more complicated. Even in this case, our calculation lead to a recurrence relation between the electronic Green's functions of the system's quantum dots, however, different than in the Hartree-Fock approximation, the electronic Green's function for a component dot relates to the electronic Green's functions in the near and next-near quantum dots. The main result of our calculation is the inclusion of several anomalous average terms, terms without which important physical effects such as the Kondo effect cannot be evaluated properly.

As an example for our general theory, we considered the case of a double-dot T-shape system consisting on a detector dot connected directly to the external leads and a side dot connected only to the detector dot and not to the external leads. In both system's dots we considered a finite on-site Coulomb interaction and we discussed in various approximations the total average electronic occupation of the system's localized levels. We found that the inclusion of the anomalous terms in the evaluation of the detector and side  dots Green's functions is very important in the evaluation of the system occupation number. Depending on the relative position of the detector dot localized level respect to the Fermi energy, different occupation regimes are possible, pointing the important role of the on-site Coulomb interaction term.

\begin{acknowledgments}
IT would like to acknowledge financial support from the Intramural Grant program at CSUF. 
\end{acknowledgments}

\begin{appendix}
\begin{widetext}
\section{Higher order correlation functions}\label{AppendixA}
In this Appendix we present the general procedure for the estimation of the four particle correlation function $\Gamma_{ij}^\sigma (\omega)=<<n_{i-\sigma}a_{i\sigma};a^\dagger_{j\sigma}>>$ (see  Eq. (\ref{gamma_ij})) using the EOM method beyond the  Hartree-Fock approximation. A similar calculation can be performed for the correlation functions in Eq. (\ref{gamma_ik}). The general equations for these functions are
\begin{eqnarray}\label{A1}
\left(\omega-\ve_\bk\right)\left<\left<n_{1-\sigma} c_{\bk\sigma};a^\dagger_{j\sigma}\right>\right>&=&
V_{\bk 1} \left<\left<n_{1-\sigma} a_{1\sigma};a^\dagger_{j\sigma}\right>\right>\nonumber\\
&&+\sum_{\bk'}V_{\bk '1}
\left[\left<\left<a^\dagger_{1-\sigma}c_{\bk '-\sigma}c_{\bk \sigma};a^\dagger_{j\sigma}\right>\right>-
\left<\left<c^\dagger_{\bk '-\sigma}a_{1-\sigma}c_{\bk\sigma};a^\dagger_{j\sigma}\right>\right>\right]\nonumber\\
&&+t_{12}\left[\left<\left<a^\dagger_{1-\sigma}a_{2-\sigma}c_{\bk\sigma};a^\dagger_{j\sigma}
\right>\right>-\left<\left<a^\dagger_{2-\sigma}a_{1-\sigma}c_{\bk\sigma};a^\dagger_{j\sigma}\right>\right>\right]\;,
\end{eqnarray}
\begin{eqnarray}\label{A2}
\left(\omega-\ve_\bk\right)\left<\left<a^\dagger_{1-\sigma}c_{\bk-\sigma}a_{1\sigma};a^\dagger_{j\sigma}\right>\right>
&=&\left<a^\dagger_{1-\sigma}c_{\bk-\sigma}\right>\delta_{1j}+
V_{\bk 1}\left<\left<n_{1-\sigma}a_{1\sigma};a^\dagger_{j\sigma}\right>\right>\nonumber\\
&&+\sum_{\bk'} V_{\bk' 1}\left[\left<\left<a^\dagger_{1-\sigma} c_{\bk-\sigma} c_{\bk'\sigma};a^\dagger_{j\sigma}\right>\right>-
\left<\left<c^\dagger_{\bk'-\sigma}c_{\bk-\sigma}a_{1\sigma};a^\dagger_{j\sigma}\right>\right>\right]\nonumber\\
&&+t_{12}\left[\left<\left<a^\dagger_{1-\sigma}c_{\bk-\sigma}a_{2\sigma};a^\dagger_{j\sigma}\right>\right>-
\left<\left<a^\dagger_{2-\sigma}c_{\bk-\sigma}a_{1\sigma};a^\dagger_{j\sigma}\right>\right>\right]\;,
\end{eqnarray}
and
\begin{eqnarray}\label{A3}
\left(\omega+\ve_\bk-2\ve_1-U_1\right)
\left<\left<c^\dagger_{\bk-\sigma}a_{1-\sigma}a_{1\sigma};a^\dagger_{j\sigma}\right>\right>&=&
\left<c^\dagger_{\bk-\sigma}a_{1-\sigma}\right>\delta_{1j}-
V_{\bk 1}\left<\left<n_{1-\sigma}a_{1\sigma};a^\dagger_{j\sigma}\right>\right>\nonumber\\
&&+\sum_{\bk'} V_{\bk' 1}\left[\left<\left<c^\dagger_{\bk-\sigma}a_{1-\sigma}c_{\bk' \sigma};
a^\dagger_{j\sigma}\right>\right>+\left<\left<c^\dagger_{\bk-\sigma}c_{\bk'-\sigma}a_{1\sigma};a^\dagger_{j\sigma}
\right>\right>\right]\nonumber\\
&&+t_{12}\left[\left<\left<c^\dagger_{\bk-\sigma}a_{1-\sigma}a_{2\sigma};a^\dagger_{j\sigma}\right>\right>
+\left<\left<c^\dagger_{\bk-\sigma}a_{2-\sigma}a_{1\sigma};a^\dagger_{j\sigma}\right>\right>\right]\;.
\end{eqnarray}
For $1\leq i \leq N-1$
\begin{eqnarray}\label{A4}
\left(\omega-\ve_{i+1}\right)&&
\left<\left<n_{i-\sigma}a_{i+1\sigma};a^\dagger_{j\sigma}\right>\right>=\nonumber\\
&&\left<n_{i-\sigma}\right>\delta_{i+1,j}+\delta_{i1}\sum_{\bk'}V_{\bk' 1}\
\left[\left<\left<a^\dagger_{i-\sigma}c_{\bk'-\sigma}a_{i+1\sigma};a^\dagger_{j\sigma}\right>\right>-
\left<\left<c^\dagger_{\bk'-\sigma}a_{i-\sigma}a_{i+1\sigma};a^\dagger_{j\sigma}\right>\right>\right]\nonumber\\
&&+t_{i,i+1}\left[\left<\left<a^\dagger_{i-\sigma}a_{i+1-\sigma}a_{i+1\sigma};a^\dagger_{j\sigma}\right>\right>
+\left<\left<n_{i-\sigma}a_{i\sigma};a^\dagger_{j\sigma}\right>\right>-
\left<\left<a^\dagger_{i+1-\sigma}a_{i-\sigma}a_{i+1\sigma};a^\dagger_{j\sigma}\right>\right>\right]\nonumber\\
&&+\Theta(i-2) t_{i-1,i}\left[
\left<\left<a^\dagger_{i-\sigma}a_{i-1-\sigma}a_{i+1\sigma};a^\dagger_{j\sigma}\right>\right>-
\left<\left<a^\dagger_{i-1-\sigma}a_{i-\sigma}a_{i+1\sigma};a^\dagger_{j\sigma}\right>\right>\right]\nonumber\\
&&+\Theta(N-2-i) t_{i+1,i+2}\left<\left<n_{i-\sigma}a_{i+2\sigma};a^\dagger_{j\sigma}\right>\right>\;,
\end{eqnarray}
\begin{eqnarray}\label{A5}
\left(\omega-\ve_{i+1}+U_i\right)&&
\left<\left<a^\dagger_{i-\sigma}a_{i+1-\sigma}a_{i\sigma};a^\dagger_{j\sigma}\right>\right>=\nonumber\\
&&\left<a^\dagger_{i-\sigma}a_{i+1-\sigma}\right>\delta_{ij}+\delta_{i1}\sum_{\bk'}
V_{\bk' 1}\left[\left<\left<a^\dagger_{1-\sigma}a_{2-\sigma}c_{\bk'\sigma};a^\dagger_{j\sigma}\right>\right>-
\left<\left<c^\dagger_{\bk'-\sigma} a_{2-\sigma}a_{1\sigma};a^\dagger_{j\sigma}\right>\right>\right]\nonumber\\
&&+t_{i,i+1}\left[
\left<\left<a^\dagger_{i-\sigma}a_{i+1-\sigma}a_{i+1\sigma};a^\dagger_{j\sigma}\right>\right>+
\left<\left<n_{i-\sigma}a_{i\sigma};a^\dagger_{j\sigma}\right>\right>-
\left<\left<n_{i+1-\sigma}a_{i\sigma};a^\dagger_{j\sigma}\right>\right>\right]\nonumber\\
&&+\Theta(i-2) t_{i-1,i}\left[\left<\left<a^\dagger_{i-\sigma}a_{i+1-\sigma}a_{i-1\sigma};a^\dagger_{j\sigma}\right>\right>-
\left<\left<a^\dagger_{i-1-\sigma}a_{i+1-\sigma}a_{i\sigma};a^\dagger_{j\sigma}\right>\right>\right]\nonumber\\
&&+\Theta(N-2-i)t_{i+1,i+2}\left<\left<a^\dagger_{i-\sigma}a_{i+2-\sigma}a_{i\sigma};a^\dagger_{j\sigma}\right>\right>
\;,
\end{eqnarray}
and
\begin{eqnarray}\label{A6}
\left(\omega-2\ve_i+\ve_{i+1}-U_i\right)&&
\left<\left<a^\dagger_{i+1-\sigma}a_{i-\sigma}a_{i\sigma};a^\dagger_{j\sigma}\right>\right>=\nonumber\\
&&\left<a^\dagger_{i+1-\sigma}a_{i-\sigma}\right>\delta_{ij}+\delta_{i1}\sum_{\bk'}V_{\bk' 1}
\left[\left<\left<a^\dagger_{i+1-\sigma}a_{i-\sigma}c_{\bk'\sigma};a^\dagger_{j\sigma}\right>\right>+
\left<\left<a^\dagger_{i+1-\sigma}c_{\bk'-\sigma}a_{i\sigma};a^\dagger_{j\sigma}\right>\right>\right]\nonumber\\
&&+t_{i,i+1}\left[\left<\left<a^\dagger_{i+1-\sigma}a_{i-\sigma}a_{i+1\sigma};a^\dagger_{j\sigma}\right>\right>+
\left<\left<n_{i+1-\sigma}a_{i\sigma};a^\dagger_{j\sigma}\right>\right>-
\left<\left<n_{i-\sigma}a_{i\sigma};a^\dagger_{j\sigma}\right>\right>\right]\nonumber\\
&&+\Theta(i-2) t_{i-1,i}\left[\left<\left<a^\dagger_{i+1-\sigma}a_{i-\sigma}a_{i-1\sigma};a^\dagger_{j\sigma}\right>\right>+
\left<\left<a^\dagger_{i+1-\sigma}a_{i-1-\sigma}a_{i\sigma};a^\dagger_{j\sigma}\right>\right>\right]\nonumber\\
&&-\Theta(N-2-i)t_{i+1,i+2}\left<\left<a^\dagger_{i+2-\sigma}a_{i-\sigma}a_{i\sigma};a^\dagger_{j\sigma}\right>\right>
\;.
\end{eqnarray}
For $2\leq i\leq N$ we have
\begin{eqnarray}\label{A7}
\left(\omega-\ve_{i-1}\right)&&
\left<\left<n_{i-\sigma}a_{i-1\sigma};a^\dagger_{j\sigma}\right>\right>=\nonumber\\
&&\left<n_{i-\sigma}\right>\delta_{i-1,j}+\delta_{i2}\sum_{\bk'} V_{\bk' 1}\left<\left<n_{i-\sigma}c_{\bk'\sigma};a^\dagger_{j\sigma}\right>\right>\nonumber\\
&&+t_{i-1,i}\left[\left<\left<n_{i-\sigma}a_{i\sigma};a^\dagger_{j\sigma}\right>\right>-
\left<\left<a^\dagger_{i-1-\sigma}a_{i-\sigma}a_{i-1\sigma};a^\dagger_{j\sigma}\right>\right>+
\left<\left<a^\dagger_{i-\sigma}a_{i-1-\sigma}a_{i-1\sigma};a^\dagger_{j\sigma}\right>\right>\right]\nonumber\\
&&+\Theta(N-1-i) t_{i,i+1} \left[
\left<\left<a^\dagger_{i-\sigma}a_{i+1-\sigma}a_{i-1\sigma};a^\dagger_{j\sigma}\right>\right>+
\left<\left<a^\dagger_{i+1-\sigma}a_{i-\sigma}a_{i-1\sigma};a^\dagger_{j\sigma}\right>\right>\right]\nonumber\\
&&+\Theta(i-3) t_{i-2,i-1} \left<\left<n_{i-\sigma}a_{i-2\sigma};a^\dagger_{j\sigma}\right>\right>\;,
\end{eqnarray}
\begin{eqnarray}\label{A8}
\left(\omega-\ve_{i-1}+U_i\right)&&
\left<\left<a^\dagger_{i-\sigma}a_{i-1-\sigma}a_{i\sigma};a^\dagger_{j\sigma}\right>\right>=\nonumber\\
&&\left<a^\dagger_{i-\sigma}a_{i-1-\sigma}\right>\delta_{ij}+\delta_{i2}\sum_{\bk'} V_{\bk' 1}
\left<\left<a^\dagger_{i-\sigma}c_{\bk'-\sigma}a_{i\sigma};\a^\dagger_{j\sigma}\right>\right>\nonumber\\
&&+ t_{i-1,i}\left[\left<\left<a^\dagger_{i-\sigma}a_{i-1-\sigma}a_{i-1\sigma};a^\dagger_{j\sigma}\right>\right>
+\left<\left<n_{i-\sigma}a_{i\sigma};a^\dagger_{j\sigma}\right>\right>-
\left<\left<n_{i-1-\sigma}a_{i\sigma};a^\dagger_{j\sigma}\right>\right>\right]\nonumber\\
&&+\Theta(N-1-i)t_{i,i+1}
\left[\left<\left<a^\dagger_{i-\sigma}a_{i-1-\sigma}a_{i+1\sigma};a^\dagger_{j\sigma}\right>\right>
-\left<\left<a^\dagger_{i+1-\sigma}a_{i-1-\sigma}a_{i\sigma};a^\dagger_{j\sigma}\right>\right>\right]\nonumber\\
&&+\Theta(i-3) t_{i-2,i-1} \left<\left<a^\dagger_{i-\sigma}a_{i-2-\sigma}a_{i\sigma};a^\dagger_{j\sigma}\right>\right>\;,
\end{eqnarray}
and
\begin{eqnarray}\label{A9}
\left(\omega-2\ve_i+\ve_{i-1}-U_i\right)&&
\left<\left<a^\dagger_{i-1-\sigma}a_{i-\sigma}a_{i\sigma};a^\dagger_{j\sigma}\right>\right>=\nonumber\\
&&\left<a^\dagger_{i-1-\sigma}a_{i-\sigma}\right>\delta_{ij}-\delta_{i2}\sum_{\bk'} V_{\bk' 1}
\left<\left<c^\dagger_{\bk'-\sigma}a_{i-\sigma}a_{i\sigma};a^\dagger_{j\sigma}\right>\right>\nonumber\\
&&+t_{i-1,i}\left[\left<\left<a^\dagger_{i-1-\sigma}a_{i-\sigma}a_{i-1\sigma};a^\dagger_{j\sigma}\right>\right>
+\left<\left<n_{i-1-\sigma}a_{i\sigma};a^\dagger_{j\sigma}\right>\right>-
\left<\left<n_{i-\sigma}a_{i\sigma};a^\dagger_{j\sigma}\right>\right>\right]\nonumber\\
&&+\Theta(N-1-i) t_{i,i+1}\left[
\left<\left<a^\dagger_{i-1-\sigma}a_{i-\sigma}a_{i+1\sigma};a^\dagger_{j\sigma}\right>\right>+
\left<\left<a^\dagger_{i-1-\sigma}a_{i+1-\sigma}a_{i\sigma};a^\dagger_{j\sigma}\right>\right>\right]\nonumber\\
&&-\Theta(i-3) t_{i-2,i-1} \left<\left<a^\dagger_{i-2-\sigma}a_{i-\sigma}a_{i\sigma};a^\dagger_{j\sigma}\right>\right>\;.
\end{eqnarray}
In all the above equations higher order correlation functions involving six particle processes were neglected. Obviously, to truncate the chain of general equations produced by the EOM method one has to make additional approximations in Eqs. (\ref{A1})-(\ref{A9}). In the following, the standard procedure we used to approximate the four operators correlation functions,  $<<A_{-\sigma}B_{-\sigma}C_\sigma;D_\sigma>>$, is to replace them by a product of the average value of two operators and the correlation function of the remaining two, $<A_{-\sigma}B_{-\sigma}>$$<<C_{\sigma};D_{\sigma}>>$. Of course, this procedure is not applied when any of the four particle correlation functions resembles one of the initial functions we try to estimate, for the particular case of Eq. (\ref{gamma_ij}), $\Gamma_{ij}^\sigma (\omega)=<<n_{i-\sigma}a_{i\sigma};a^\dagger_{j\sigma}>>$. Based on this approximation we rewrite Eq. (\ref{gamma_ij}) as
\begin{eqnarray}\label{A_gamma_ij}
&&\left[\omega-\ve_i-U_i-\delta_{i1}\left(2\Sigma_\bk^0(\omega)+\Sigma_\bk^1(\omega)\right)-
\Theta(N-1-i) t^2_{i,i+1} M_{i,i+1}(\omega) - \Theta(i-2) t^2_{i-1,i} M_{i,i-1}(\omega)\right]\Gamma_{ij}^\sigma(\omega)=\nonumber\\
&&\left<n_{i-\sigma}\right>\left[\delta_{ij}+\delta_{i2}\Theta(i-2) t_{i-1,i} \f{\sum_\bk' V_{\bk'1}G_{\bk'j}^\sigma(\omega)}{\omega-\ve_i}
+\Theta(N-1-i)\f{t_{i,i+1}}{\omega-\ve_{i+1}}\delta_{i+1,j}+\Theta(i-2)\f{t_{i-1,i}}{\omega-\ve_{i-1}}\delta_{i-1,j}
\right.\nonumber\\
&&\left.\hspace{1.1cm}+\Theta(N-2-i)\f{t_{i+1,i+2}}{\omega-\ve_{i+1}}G_{i+2,j}^\sigma(\omega)+\Theta(i-3)\f{t_{i-2,i-1}}{\omega-\ve_{i-1}}
G_{i-2,j}^\sigma(\omega)\right]\nonumber\\
&&-G_{ij}^\sigma(\omega)\left[\delta_{i1}\Sigma_\bk^2(\omega)+\Theta(N-1-i)t^2_{i,i+1}\left<n_{i+1-\sigma}\right>
P_{i,i+1}(\omega)-\Theta(i-2) t^2_{i-1,i} \left<n_{i-1-\sigma}\right>P_{i,i-1}(\omega)\right]\nonumber\\
&&+\delta_{i1} A_{ij}^\sigma(\omega) +\Theta(N-1-i) t_{i,i+1} B_{ij}^\sigma(\omega)+\Theta(i-2) t_{i-1,i} C_{ij}^\sigma(\omega)\;,
\end{eqnarray}
where for simplicity we introduced the notations
\begin{equation}
M_{ij}(\omega)=\f{1}{\omega-\ve_j}+\f{1}{\omega-\ve_{j}+U_i}+\f{1}{\omega-2\ve_i+\ve_j-U_i}\;,
\end{equation}
and
\begin{equation}
P_{ij}(\omega)=\f{1}{\omega-\ve_{j}+U_i}+\f{1}{\omega-2\ve_i+\ve_j-U_i}\;,
\end{equation}
\begin{eqnarray}
&&A_{ij}^\sigma(\omega)=\left(\delta_{ij}+\sum_{\bk'} V_{\bk'1} G_{\bk'j}^\sigma(\omega)+t_{12} G_{2j}^\sigma(\omega)\right) \sum_\bk V_{\bk1}\left(\f{\left<a^\dagger_{1-\sigma} c_{\bk-\sigma}\right>}{\omega-\ve_\bk}-\f{\left<c^\dagger_{\bk-\sigma} a_{1-\sigma}\right>}{\omega+\ve_\bk-2\ve_1-U_1}\right)\nonumber\\
&&+t_{12}\sum_\bk V_{\bk1}\left(\f{\left<a^\dagger_{2-\sigma} c_{\bk-\sigma}\right>}{\omega-\ve_\bk}-\f{\left<c^\dagger_{\bk-\sigma} a_{2-\sigma}\right>}{\omega+\ve_\bk-2\ve_1-U_1}\right)\nonumber\\
&&+\sum_\bk V_{\bk1}G_{\bk j}^\sigma(\omega)
\left[\f{\sum_{\bk'} V_{\bk'1}\left(\left<a^\dagger_{1-\sigma} c_{\bk'-\sigma}\right>-
\left<c^\dagger_{\bk'-\sigma}a_{1-\sigma} \right>\right)}{\omega-\ve_\bk}+\f{t_{12}}{\omega-\ve_\bk}
\left(\left<a^\dagger_{1-\sigma}a_{2-\sigma}\right>-\left<a^\dagger_{2-\sigma}a_{1-\sigma}\right>\right)\right],\;\;
\end{eqnarray}
\begin{eqnarray}
&&B_{ij}^\sigma(\omega)=\left(\delta_{ij}
+\delta_{i1}\sum_{\bk'} V_{\bk'1} G_{\bk'j}^\sigma(\omega)+\Theta(i-2) t_{i-1,i} G_{i-2,j}^\sigma(\omega)\right)\left[\f{\left<a^\dagger_{i-\sigma} a_{i+1-\sigma}\right>}{\omega-\ve_{i+1}+U_i}-\f{\left<a^\dagger_{i+1-\sigma} a_{i-\sigma}\right>}{\omega-2\ve_i+\ve_{i+1}-U_i}\right]\nonumber\\
&&+G_{i+1,j}^\sigma(\omega)\left[\delta_{i1}\f{\sum_{\bk'} V_{\bk'1}\left(\left<a^\dagger_{i-\sigma} c_{\bk'-\sigma}\right>-\left<c^\dagger_{\bk'-\sigma} a_{i-\sigma}\right>\right)}{\omega-\ve_{i+1}}+
\Theta(i-2) t_{i-1,i}\f{\left<a^\dagger_{i-\sigma}a_{i-1-\sigma}\right>-
\left<a^\dagger_{i-1-\sigma} a_{i-\sigma}\right>}{\omega-\ve_{i+1}}\right.\nonumber\\
&&\hspace{1.6cm}\left.+t_{i,i+1}\left<a^\dagger_{i-\sigma} a_{i+1-\sigma}\right>\left(\f{1}{\omega-\ve_{i+1}}+
\f{1}{\omega-\ve_{i+1}+U_i}\right)\right.\nonumber\\
&&\hspace{1.6cm}\left.-t_{i,i+1}\left<a^\dagger_{i+1-\sigma}a_{i-\sigma}\right>
\left(\f{1}{\omega-\ve_{i+1}}+\f{1}{\omega-2\ve_i+\ve_{i+1}-U_i}\right)\right]\nonumber\\
&&-G_{ij}^\sigma(\omega)\left[\delta_{i1}\sum_{\bk'} V_{\bk'1}\left(\f{\left<c^\dagger_{\bk'-\sigma}a_{2-\sigma}\right>}{\omega-\ve_{i+1}+U_i}+
\f{\left<a^\dagger_{2-\sigma} c_{\bk'-\sigma}\right>}{\omega-2\ve_i+\ve_{i+1}-U_i}\right)\right.\nonumber\\
&&\hspace{1.5cm}+\left.t_{i-1,i}\Theta(i-2)\left(\f{\left<a^\dagger_{i-1-\sigma} a_{i+1-\sigma}\right>}{\omega-\ve_{i+1}+U_i}+\f{\left<a^\dagger_{i+1-\sigma} a_{i-1-\sigma}\right>}{\omega-2\ve_i+\ve_{i+1}-U_i}\right)\right.\nonumber\\
&&\hspace{1.5cm}+\left.t_{i+1,i+2}\Theta(N-2-i)
\left(\f{\left<a^\dagger_{i-\sigma} a_{i+2-\sigma}\right>}{\omega-\ve_{i+1}+U_i}+
\f{\left<a^\dagger_{i+2-\sigma} a_{i-\sigma}\right>}{\omega-2\ve_i+\ve_{i+1}-U_i}\right)\right]\;,
\end{eqnarray}
and
\begin{eqnarray}
C_{ij}^\sigma(\omega)&=&\left(\delta_{ij}+\Theta(N-1-i) t_{i,i+1}\right)
\left[\f{\left<a^\dagger_{i-\sigma} a_{i-1-\sigma}\right>}{\omega-\ve_{i-1}+U_i}-
\f{\left<a^\dagger_{i-1-\sigma} a_{i-\sigma}\right>}{\omega-2\ve_i+\ve_{i-1}-U_i}\right]\nonumber\\
&&+G_{ij}^\sigma(\omega)\left[\delta_{i2}\sum_{\bk'} V_{\bk'1}\left(
\f{\left<a^\dagger_{i-\sigma} c_{\bk'-\sigma}\right>}{\omega-\ve_{i-1}+U_i}+
\f{\left<c^\dagger_{\bk'-\sigma} a_{i-\sigma}\right>}{\omega-2\ve_i+\ve_{i-1}-U_i}\right)\right.\nonumber\\
&&\left.\hspace{1.5cm}-\Theta(N-1-i) t_{i,i+1}\left(\f{\left<a^\dagger_{i+1-\sigma} a_{i-1-\sigma}\right>}{\omega-\ve_{i-1}+U_i}-
\f{\left<a^\dagger_{i-1-\sigma} a_{i+1-\sigma}\right>}{\omega-2\ve_i+\ve_{i+1}-U_i}\right)\right.\nonumber\\
&&\left.\hspace{1.5cm}+\Theta(i-3) t_{i-2,i-1} \left(\f{\left<a^\dagger_{i-\sigma} a_{i-2-\sigma}\right>}{\omega-\ve_{i-1}+U_i}+
\f{\left<a^\dagger_{i-2-\sigma} a_{i-\sigma}\right>}{\omega-2\ve_i+\ve_{i-1}+U_i}\right)\right]\nonumber\\
&&+G_{i-1,j}^\sigma(\omega)\left[t_{i-1,i}\left<a^\dagger_{i-\sigma} a_{i-1-\sigma}\right>\left(\f{1}{\omega-\ve_{i-1}}+
\f{1}{\omega-\ve_{i-1}+U_i}\right)\right.\nonumber\\
&&\hspace{1.6cm}\left.-t_{i-1,i}\left<a^\dagger_{i-1-\sigma}a_{i-\sigma}\right>
\left(\f{1}{\omega-\ve_{i-1}}+\f{1}{\omega-2\ve_i+\ve_{i-1}-U_i}\right)\right.\nonumber\\
&&\hspace{1.6cm}\left.+\Theta(N-1-i)\f{t_{i,i+1}}{\omega-\ve_{i-1}}\left(
\left<a^\dagger_{i-\sigma} a_{i+1-\sigma}\right>+\left<a^\dagger_{i+1-\sigma} a_{i-\sigma}\right>\right)\right]\;.
\end{eqnarray}

Additionally, similar to Lacroix \cite{lacroix} we introduced two terms related to the interaction between free electrons in the leads and the localized electrons in the detector quantum dot: 
\begin{equation}\label{sigma1}
\Sigma_\bk^1(\omega)=\sum_\bk\f{V^2_{\bk1}}{\omega+\ve_\bk-2\ve_1-U_1}\;,
\end{equation}
\begin{equation}\label{sigma2}
\Sigma_\bk^2(\omega)=\sum_\bk V^2_{\bk1}  \left(\f{\sum_{\bk'}\left<c^\dagger_{\bk'-\sigma}c_{\bk-\sigma}\right>}{\omega-\ve_\bk}+
\f{\sum_{\bk'}\left<c^\dagger_{\bk-\sigma}c_{\bk'-\sigma}\right>}{\omega+\ve_\bk-2\ve_1-U_1}\right)\;.
\end{equation}
As a general remark all the above terms introduce anomalous correlations between electrons in different quantum dots or between electrons in the leads and in the detector quantum dot. Such terms are usually neglected; however, it was proved that they are of main interest when physics associated with the Kondo effect is discussed \cite{lacroix}. Consider for example the simple situation in which the T-shape system consists only in a single quantum dot ($t_{12}=0$) and focus on the anomalous terms introduced by Eq. (\ref{sigma2}), i.e.,  $\left<c^\dagger_{\bk'-\sigma}c_{\bk-\sigma}\right>$ and $\left<c^\dagger_{\bk-\sigma}c_{\bk'-\sigma}\right>$. The simplest approximation will be to replace these terms with $\delta_{\bk\bk'} f(\ve_\bk)$, although a more careful analysis shows that
\begin{equation}
\left<c^\dagger_{\bk'-\sigma}c_{\bk-\sigma}\right>=-\f{1}{\pi}\int_{-\infty}^\infty f(\omega)\; \textrm{Im}\left<\left< c_{\bk-\sigma}; c^\dagger_{\bk'-\sigma}\right>\right>,
\end{equation}
with
\begin{equation}
\left<\left< c_{\bk-\sigma}; c^\dagger_{\bk'-\sigma}\right>\right>=\f{\delta_{\bk\bk'}}{\omega-\ve_\bk}+\f{V_{\bk 1} V_{\bk'1}}{(\omega-\ve_\bk)(\omega-\ve_{\bk'})}\; G^{-\sigma}_{11} (\omega)\;.
\end{equation}
Under this approximation, if we consider terms up to the quadratic order in $V_{\bk1}$ we find
\begin{equation}
\Sigma^2_\bk (\omega)\simeq \f{\Delta}{\pi}\left[\Psi\left(\f{1}{2}+\f{\omega-2\ve_1-U_1}{2\pi i T}\right)-\Psi\left(\f{1}{2}+\f{\omega}{2\pi i T}\right)-i\pi\right]\;.
\end{equation}
\end{widetext}

\end{appendix}

\end{document}